\documentclass{aa}
\usepackage{graphicx}
\begin{document}

\title{\bf Hot Flashers and He Dwarfs in Galactic Globulars}

\author{M.Castellani \inst{1} \and V. Castellani \inst{1} 
\thanks{It is with deep sorrow that we announce the death of Vittorio 
Castellani, who passed away on May 20, 2006.}
\and P.G. Prada Moroni \inst{2,3} }
\offprints {M. Castellani, \email {m.castellani@mporzio.astro.it}}
\institute{INAF, Osservatorio Astronomico di Roma,
Via Frascati 33, 00040 Monteporzio Catone, Italy
\and
Dipartimento di Fisica, Universit\`a di Pisa,
Largo Pontecorvo 3, 56127 Pisa, Italy
\and
INFN, Sezione di Pisa, Largo Pontecorvo 3,
56127 Pisa, Italy
}

\date{Received ; accepted }

%\maketitle
%

\abstract
% context heading
{We  revisit the evolutionary scenario for Hot Flasher low-mass
structures, where mass loss delays the He flash till the initial
phases of their White Dwarf cooling sequence.}
% aims heading
{Our aim has been to test the theoretical results vis-a-vis different
assumptions about the efficiency of mass loss.}
% method heading
{To this purpose, we present evolutionary models covering a fine grid
of masses, as obtained assuming a single episode of mass loss in a
Red Giant model of 0.86 $M_{\odot}$ with Z=0.0015.}
% results heading
{We find a reasonable agreement with previous evolutionary
investigations, showing that for the given metallicity late Hot
Flashers are predicted to cover the mass range M=0.4975 to M= 0.4845
($\pm$0.0005) $M_{\odot}$, all models igniting the He-flash with a
mass of the H-rich envelope as given by $M_e$=0.00050
$\pm$0.00002 $M_{\odot}$. The ignition mechanism is discussed in some
details, showing the occurrence of a bifurcation in the evolutionary
history of stellar structures at the lower mass limit for He
ignition. Below such a critical mass, the structures miss the He
ignition, cooling down as a Hot Flasher-Manqu\'e He White Dwarf.  We
predict that these structures will cool down, reaching the
luminosity $logL/L_{\odot}$=-1 in a time at the least five times longer than
the corresponding cooling time of a normal CO White Dwarf.}
% conclusions heading
{On very general grounds, one expects that old stellar clusters with
a sizeable population of Hot Flasher should likely produce at least
a similar amount of slow-cooling  He White Dwarfs. According to this
result, in a cluster where 20\% of Red Giants escape the He burning
phase, one expects roughly twice as White Dwarfs than in a normal
cluster where all Red Giants undergo their He flash}


   \maketitle

%__________________________________________________

\section{Introduction}

Over the last decades the evolution of low-mass stellar structures
has been the subject of a large amount  of investigations, aimed at
constraining the evolutionary status of stars in  old stellar
systems, such as Galactic Globular Clusters. Since long time we know
that present  Globular Cluster stars are expected to leave their
Main Sequence to climb along the Red Giant Branch (RGB) till the
onset of the He-flash. After the phase of central (Horizontal
Branch) and shell (Asymptotic Giant Branch) He burning phases, they
will eventually cool down under the form of Carbon-Oxygen  (CO)
White Dwarfs (WDs).

In this context,  the occurrence of extended "Blue Tails" in the
Horizontal Branches (HB) of several Galactic Globulars has already
been understood in terms of RGB structures which have lost the large
majority of their H-rich  envelope before igniting He to become HB
stars. Castellani \& Castellani (1993; but see also Castellani,
Degl'Innocenti \& Pulone 1995) found that, for extreme mass loss,
there are stellar models which fail to ignite He at the tip of the
RGB, but undergo a late He-flash during the contraction toward their
He-WD structure or in the early stages of the WD cooling sequence.
Similar structures are now known in the literature as "Hot
He-Flashers" (HFs). Even larger mass loss will prevent the He ignition,
definitively producing He White Dwarfs.

Hot Flashers have been extensively investigated by several
authors. D'Cruz et al. (1996) made use of Reimers (1975, 1977)
formula for mass loss, taking  the efficiency parameter $\eta_R$ as
a free parameter to explore the range of mass-loss producing HFs for
selected assumptions about the star metallicity. Sweigart (1997)
discovered that when the He-flash occurs along the WD cooling
sequence ("late" HFs), then convection  is expected to reach the
H-rich envelope, enhancing  He and Carbon abundances in the stellar
atmosphere and driving strong H-flashes.  Brown et al. (2001)
adopted again the Reimers formalism to explore the occurrence of
late HF for the metal abundance Z=0.0015, in connection with
observational evidence for extremely hot HB stars in the Galactic
globular NGC2808. Quantitative estimates of the mixing driven by the
He-flash have been finally presented by Cassisi et al. (2003), who
were able, for the first time, to follow in detail the growth of
such an instability in late HFs.

In this paper we revisit the  HF theoretical scenario but
adopting different assumptions concerning the mass-loss
mechanism. On this basis we will present and discuss  new
evolutionary results, focusing the attention on the stellar
structures marking the transition between late HF and bona fide He
WD.

\section{The models}

All the papers  quoted in the previous section have investigated the
occurrence of HF using the Reimers mass-loss parameter $\eta_R$ as a
free parameter to govern the efficiency and, thus, the amount of
mass loss. However, among these investigations there are subtle
differences. As a matter of fact Castellani \& Castellani (1993)
took into account  mass loss till the onset of the He-flash, whereas D'Cruz
et al. (1996) neglected mass loss when the mass of the H-rich
stellar envelope reached the value $M_e=10^{-3} M_{\odot}$.
Brown et al (2001)
also stopped the mass loss, but when the structure moved away from
the RGB by $\Delta logT_e[K] =0.1$.

Here we notice that, at least in principle, these differences can
have sizeable consequences on the final structures. Data in Table 4
of Castellani \& Castellani (1993) disclose that, according to
Reimers's formulation, after leaving the RGB a HF model  is expected
to loose an amount of mass of the same order of magnitude of the
value of $M_e$ at the onset of the flash ($M_e^f$). In turn,
theoretical predictions on $M_e^f$ are at the basis of   relevant
observational constraints, since the minimum value of $M_e^f$
governs the maximum effective temperature that can be reached by
normal Zero Age Horizontal Branch (ZAHB) models: the maximum
effective temperature increases when $M_e^f$ decreases (see, e.g.,
the discussion in Castellani, Degl'Innocenti \& Pulone 1995). It
appears thus of obvious relevance to investigate in detail
theoretical predictions on such a critical  issue.

However, in spite of the different assumptions about mass loss, all
the investigations we are referring to find rather similar values
for $M_e^f$. This interesting evidence is supplemented by the result
by Brown et al. (2001, but see also D'Cruz et al. 1996) who found
that, for a given original chemical composition  {\it  late HFs have
all  exactly the same value of $M_e^f$}. Such a behavior  suggest
that $M_e$ could be the parameter governing the onset of the flash
in HF structures, independently of any assumptions about the
mechanism and the efficiency of mass loss. Such a suggestion can be
supported by inspection of evolutionary data for HF structures, as
given in the already quoted Table 4 in Castellani \& Castellani
(1993). Stars leaving the RGB  are still supported by CNO H-shell
burning. Only when approaching the final ("critical") value of $M_e$
the CNO burning starts decreasing and the stellar core experiences
the final contraction leading the structure either to a late
He-flash or to the final cooling as a He-WD. Here we  suggest to
regard the quoted time sequence as an evidence that in late HFs
there is a critical minimum $M_e^f$ value supporting H-shell
burning. When this minimum is reached, the H-shell switches-off,
causing the core contraction and the switch-on of the He-flash. As
well known, the contrary occurs in normal RGB structures
experiencing the He-flash.
%==========================TABLE 1=======================

\begin{table}
\caption{ Selected physical quantities for stellar structures at the
He flash ignition as a function of the mass after the episode of
mass loss. Masses and luminosities are in solar units. }

\begin{center}
\begin{tabular}{l r r r r }
\hline
$ M$ & $ logL^f $ & $ logT_e^f $ & $ M_c^f $  & $M_e^f$  \\
\hline
{\it   RG Flash}\\
0.5301 & 3.3915 & 3.6221 & 0.49755 & 0.03255  \\
0.5200 & 3.3905 & 3.6287 & 0.49735 & 0.02265\\
0.5100 & 3.3885 & 3.6388 & 0.49715 &  0.01285\\
0.5028 & 3.3856 & 3.6580 & 0.49663 & 0.00617\\
\hline
{\it HF-Transition}\\
0.5000 & 3.1639 & 4.8936& 0.49914 &  0.00086\\
0.4980 & 3.1745 & 4.9968 & 0.49723 &   0.00077\\
\hline
{\it Late HF} \\
0.4970 & 0.5872 & 4.6490  & 0.49652&    0.00048\\
0.4950 & 0.1837 &4.5160 & 0.49451 &  0.00049\\
0.4910 & -0.1295 & 4.4438 & 0.49050& 0.00050 \\
0.4901 &  -0.1411 & 4.4388  & 0.48960 & 0.00050  \\
0.4890 & -0.2653 & 4.4122 & 0.48848 & 0.00052  \\
0.4880 & -0.2933 & 4.4038 & 0.48748 & 0.00052  \\
0.4870 & -0.3731 & 4.3914 & 0.48649 & 0.00051  \\
0.4860 & -0.4066 & 4.3907 & 0.48549& 0.00051 \\
0.4850 & -0.4623 & 4.3917 & 0.48448 & 0.00052 \\

\hline \hline
\end{tabular}
\end{center}
\end{table}
%==========================================================

%%%%%%%%%%%%%%%%%%%%%%%%%%%  FIGURA
\begin{figure}
\centering
	\includegraphics[width=8cm]{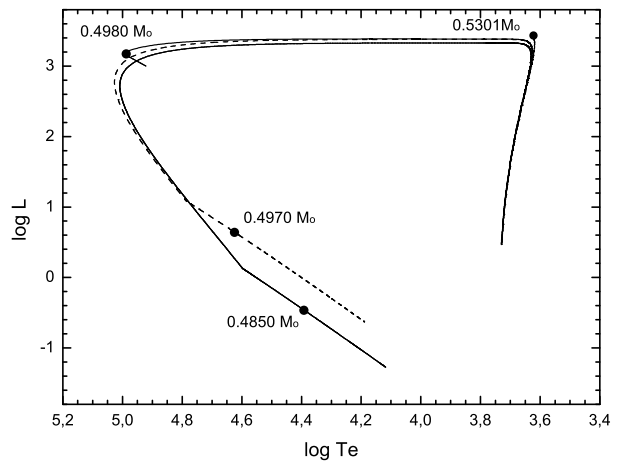}
\caption {The evolutionary paths of selected models, for the labeled
values of the total stellar mass. Dots show the location of the
He-flash as given in Table 1 (see text for more details). {\it L} is in
fraction of $L_{\odot}$, {\it Te} is in [K]. }
 \label{f:figHR}
\end{figure}

%%%%%%%%%%%%%%%%%%%%%%%%%%%

To investigate this scenario we decided to get rid of any interplay
between the late evolution of HF models and mass-loss rates,
producing suitable HF progenitors by peeling-off a model during an
early stage of his RG evolution and thus following the evolution of
the structure assuming no further mass loss. One may notice that a
similar procedure has been already adopted by Serenelli et al.
(2002).  All models have been computed assuming an initial helium
abundance Y=0.23 and a solar scaled heavy-element abundance
Z=0.0015, i.e., the same original composition adopted by Brown et
al. (2001). If not otherwise stated, in the following we will refer
to HF progenitors as obtained by applying suitable mass loss in a
0.86 M$_{\odot}$ RG structure, when $logL/L_{\odot}$=0.49 and mass
of the He core $M_c=0.16M_{\odot}$.

Table 1 gives selected physical quantities for a sample of models
experiencing the He-flash. Left to right one finds the mass of the
evolving star and, in the order, the luminosity, the effective
temperature, the mass of the He-core and the mass of the H-rich
envelope at the He flash ignition. As already known, one finds that
decreasing the stellar mass the ignition of the He flash moves from
the RG branch first to the luminous structures crossing the HR
diagram toward the WD cooling sequence (transition HF) and then to
models igniting He along the cooling sequence itself, down to
$logL/L_{\odot}\sim -0.5$ (late HF: LHF). After the He-flash, all the LHF
structures have been found to experience the mixing episode and the
explosive burning of H already discussed by Brown et al. (2001) and
Cassisi et al. (2003).

The development of the He-flash in HF structures deserves some
further comments. As usual (see, e.g., Sweigart \& Gross  1976), for
structures experiencing the He-flash either at the RG tip (RG Flash)
or during the crossing of the HR diagram toward the cooling sequence,
data in Table 1  refer to the model where the output of the
3$\alpha$ reaction has reached 100 $L_{\odot}$. However, for LHFs
the identification of the He "flashing" model is less
straightforward, since the He ignition has a rather long
evolutionary history. Taking as an example the 0.4910 $M_{\odot}$
model, the threshold of $L_{3\alpha} = 100 L_{\odot}$ is reached
along the cooling sequence at $logL/L_{\odot}$=0.52592. However, the structure
keeps cooling, whereas the efficiency of He burning increases, and
the maximum of the 3$\alpha$ production ($L_{3\alpha} = 9.2*10^9
L_{\odot}$) is reached only when $logL/L_{\odot}$= -0.1295.

According to such an evidence, we decided to list in Table 1 data
for LHF models at the first maximum of the He-flash. Fig.
\ref{f:figHR}  reports the evolutionary paths of selected HF models.
As shown in that figure, in LHF structures the onset of the He-flash
is witnessed by a clear discontinuity in the slope of the cooling
track. However, the flash attains its maximum after a not negligible
evolution along the new cooling curve. As already found by Cassisi
et al. (2003), after the flash the structure keeps cooling, until
the mixing event causes the H-flashes which drive the star toward
their stage of quiescent central He burning.

%%%%%%%%%%%%%%%%%%%%%%%%%%%  FIGURA 2
\begin{figure}
\centering
\includegraphics[width=8cm]{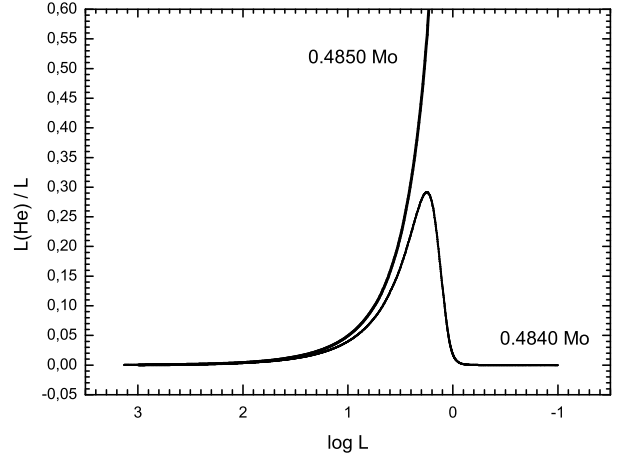}
\caption {The He burning contribution to the star luminosity for the
two cooling structures at the transition between Hot Flashers and He
White Dwarfs. {\it L} is in
fraction of $L_{\odot}$}.
 \label{f:fig2}
\end{figure}

%%%%%%%%%%%%%%%%%%%%%%%%%%%
%%%%%%%%%%%%%%%%%%%%%%%%%%%  FIGURA 3
\begin{figure}
\centering
\includegraphics[width=8cm]{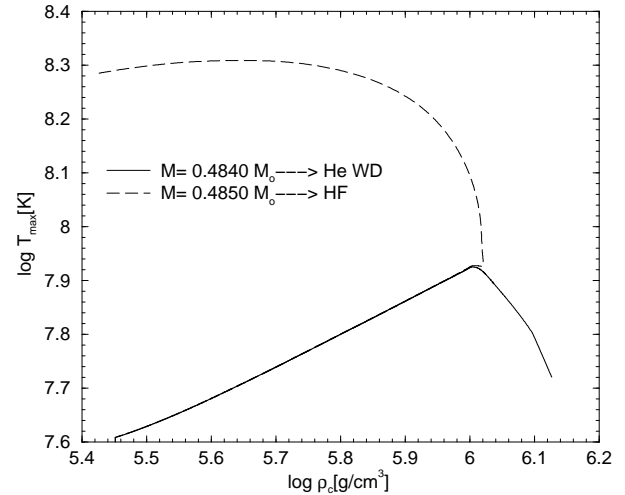}
\caption {The run of the maximum (off center) temperature versus the
central density for the two cooling structures at the transition
between Hot Flashers and He White Dwarfs. }
 \label{f:fig3}
\end{figure}

%%%%%%%%%%%%%%%%%%%%%%%%%%%

Inspection of the evolutionary results reveals the reason why below
the critical mass $M= 0.485 M_{\odot}$ the stars miss the He flash:
decreasing the mass of a Hot Flasher the contraction driving the
efficiency of 3$\alpha$ reactions starts in more advanced phases
along the cooling sequence, progressively approaching the region
where structures are strongly affected by neutrino emission. Fig.
\ref{f:fig2} shows the  contribution of 3$\alpha$ reactions to the
star luminosity for the two cooling models of  $M= 0.4850$ and
$0.4840 M_{\odot}$, i.e., just above and below the lower limit for
He ignition, at the transition between Hot Flashers and He White
Dwarfs. One finds that just below the critical mass the structure
starts attempting the final He ignition, but the neutrino emission
soon overcomes the output of nuclear energy, cooling down the
stellar interior, halting the ignition mechanism. Fig.\ref{f:fig3}
shows the maximum off center temperature as a function  of central
density for the same two models. As already found when discussing
the critical stellar mass for carbon ignition (see Fig. 4 in
Castellani et al. 2003), even in the  case of He ignition one finds
the clear evidence for the predictable  "bifurcation" in the
evolution of the physical conditions in the stellar interiors.

Data in Table 1 deserve several comments. One finds that all the LHF
reach the flashing phase  with a quite similar envelope mass, namely
$M_{e} =0.00051 \pm 0.00001 M_{\odot}$. Such an evidence reinforce the
suggestion that $M_{e}$ should be the parameter governing the onset
of the late He-flashes. At the same time, this result  confirms the
evolutionary scenario presented by Brown et al (2001). Only the mass
of the LHF envelopes is in our models slightly smaller, 0.00051
against 0.0006 $M_{\odot}$, perhaps for small differences in the
input physics (but see also Cassisi et al. (2003) who found $M_e =
0.00055 M_{\odot}$). Here we notice that the suggested connection between the
H-burning switch-off and the onset of late He-flashes gives a
natural explanation of the dependence of $M_e^f$ on metallicity
(D'Cruz et al. 1996), for which an increase in the metallicity of HF
structures causes a  decrease in  the final pre-flash $M_e$. This
appears indeed in agreement with the often reported evidence that an
increase in the metallicity decreases the minimum mass of the
envelope supporting H-shell burning (see, e.g., Castellani et al.
1994).

\section{Hot Flashers and He White Dwarfs}

\begin{figure}
\centering
\includegraphics[width=8cm]{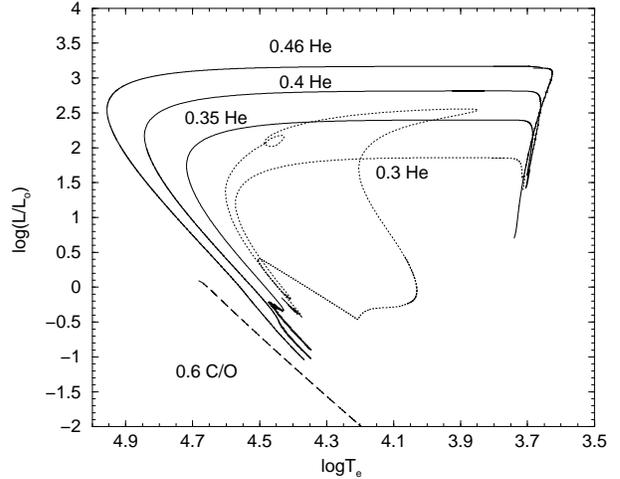}
\caption {Evolutionary paths of selected HF-Manqu\'e He WD
 models for the labelled
values of the stellar mass. The cooling sequence of a $0.6
M_{\odot}$ CO WD is reported for comparison.}
 \label{f:fig1}
\end{figure}

According to our computations, one finds a lower mass limit for the
He-flash $M = 0.4850 M_{\odot}$: a decrease of  this mass by only
$0.001 M_{\odot}$ implies that the stellar structure does not
succeed in igniting He, definitely cooling down as a He WD.
Fig.\ref{f:fig1} shows the evolutionary paths for  some of these He
WD models. As already discussed in Castellani et al. (1994), one
finds that, below the critical mass for the onset of the He flash,
the more massive WDs  are predicted to  cool down quietly at least
down to $logL/L_{\odot}$=-1.0. However, when the WD mass is decreased, the
models show  below $logL/L_{\odot}\sim$ 0 the progressive contribution of the
H-reignition which in the $0.3 M_{\odot}$ eventually  drives the
ignition of strong CNO flashes.

%%%%%%%%%%%%%%%%%%%%%%%%%%%  FIGURA 1

%%%%%%%%%%%%%%%%%%%%%%%%%%%

One might notice that the predicted LHF cover a restricted but not
negligible range of masses, namely $\Delta M \sim 0.012 M_{\odot}$,
against the mass dispersion by $\sim 0.02 M_{\odot}$ often taken as
representative of normal HB (Rood 1973). Thus the production of a
substantial amount of HF does not require a too  fine  tuning of the
mass loss mechanism. This supports the suggestion by Brown et al.
(2001) for which LHF could be at the origin of the hottest group of
HB observed in the globular NGC2808, as well as  of similar
structures in $\omega$ Cen (Moehler at al. 2004). As already
discussed by D'Cruz et al. (1996), there are no physical basis for
predicting the efficiency of mass loss. However, one can hardly
believe in a distribution of mass loss producing LHFs but no WDs. On
the contrary, the occurrence of HFs likely implies the occurrence of
at least a similar amount of stars which miss the He flash, cooling
down as He WDs.

According to their larger heat capacity, we know that He WD evolve
more slowly than CO WD do, the cooling low depending -for each given
WD mass- on the mass of the H rich envelope as well as on the
efficiency of element diffusion mechanisms. Having established
within $10^{-3}M_{\odot}$ the minimum HF mass, we have
simultaneously fixed, within the same uncertainty, the maximum mass
allowed for He dwarfs together with a firm predictions about the
mass of the H-rich envelopes in the various cooling structures. It
is obvious interesting to investigate the cooling law for these
structures in the range of masses just below the HF interval. This
appears of particular relevance since current investigations have
already detected an unsuspected large amount of Hot Flashers
candidate in clusters like NGC2808 (Castellani et al. 2006) and
$\omega$ Cen (D'Cruz et al. 2002), but also a huge amount of WDs in
the latter cluster (Monelli et al. 2005).

%==========================TABLE 2=======================

\begin{table}
\caption{ Selected physical quantities for stellar structures
cooling down as He White Dwarfs. Masses and luminosities are in
solar units, cooling times t in Myr. }

\begin{center}
\begin{tabular}{r r r r r }
\hline
$ M$ & $ logL$ & $ t  $ & $ M_c $  & $M_e$  \\
\hline

0.4840 & 0.0000 & 3.8 & 0.48342   & 0.00058\\
       & -1.0000 & 92.9 & 0.48352   & 0.00048\\
0.4600 & 0.0000 & 2.1 & 0.45642 & 0.00358  \\
       & -1.0000 &  111.4 & 0.45653 & 0.00347\\
0.4000 & 0.0000& 0.8 & 0.39834   & 0.00166 \\
       & -1.0000& 174.2 & 0.39859   & 0.00141 \\
\hline \hline
\end{tabular}
\end{center}
\end{table}
%==========================================================

Table 2 gives the cooling times for selected HF-Manqu\'e He WDs at
the two luminosities $logL/L_{\odot}$ = 0.0 and -1.0 together with
the corresponding values of the masses of the He-core and of the
H-rich envelope. Evolutionary times are computed starting from the
models reaching the maximum effective temperature after leaving the
RG branch. Inspection of evolutionary times in Table 2 reveals that
at the limit $logL/L_{\odot}$=-1 HF-Manqu\'e He WD should have a minimum
cooling age of the order of 90 Myr, which increases when the mass
decreases. Thus, above the quoted  luminosity limit HF-Manqu\'e He
WD have a lifetime of the same order of magnitude or larger than
typical HB stars.

\begin{figure}
\centering
\includegraphics[width=8cm]{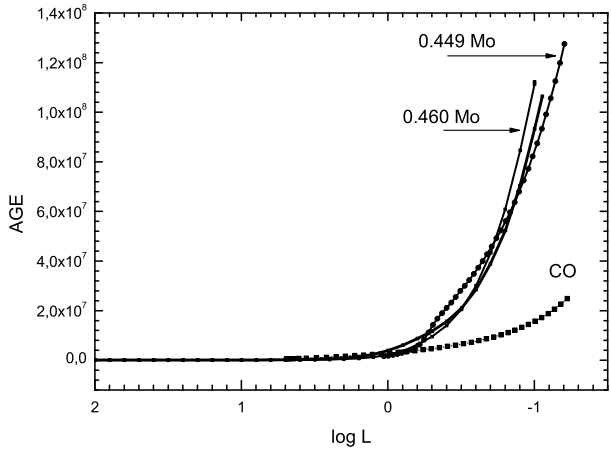}
\caption {The cooling ages versus the WD luminosity for our most
massive He WD model ($0.4840 M_{\odot}$: heavy line) compared with
the less massive model $0.4600 M_{\odot}$, the  $0.449 M_{\odot}$
with element diffusion (Serenelli et al. 2002) and a $0.5 M_{\odot}$
CO WD (Prada Moroni \& Straniero 2002). {\it L} is in fraction of $L_{\odot}$
and {\it AGE} is in years.}
 \label{f:fig4}
\end{figure}

However, one must notice that models in Table 2 do not account for
the effect of element diffusion. According to current estimates, the
main effect of diffusion in He WDs is to drive the occurrence of
strong CNO flashes which can deeply affect the age-luminosity
relation (Althaus, Serenelli \& Benvenuto 2001a, 2001b). However,
and luckily enough, even if diffusion is taken into account He WD
models just below the LHF mass limit seem to escape such an
instability, cooling quietly down toward their fainter evolutionary
phases (Serenelli et al.  2002). Thus, element diffusion should have
only minor effects. This is confirmed by data in Fig. \ref{f:fig4},
where we compare the age-luminosity relation for two selected models
from our sample with a similar model ($M=0.449 M_{\odot}$) presented
by Serenelli et al (2002) for Z=0.001 and with diffusion taken into
account. As a whole, one may conclude that age given in Table 2
should give, at least, a reasonable order of magnitude for WD ages.

Comparison with the age-luminosity relation for a CO WD  ($M=0.5
M_{\odot}$), as given in the same figure from Prada Moroni \&
Straniero (2002), gives an impressive evidence of the large
predicted differences between HF-Manqu\'e He WD and CO WD,
supporting some relevant considerations. One can easily predict,
e.g., that in a cluster where 20\% of RG escape the He burning
phase, above $logL/L_{\odot}$=-1 one expects roughly twice as WD  than in a
normal cluster where all RG undergo their He flash. It follows that
WD counts above the quoted luminosity limit can give relevant
information on the abundance of HF-Manqu\'e He WDs.

\section{Summary and conclusions}

%%%%%%%%%%%%%%%%%%%%%%%%%%%  FIGURA

%%%%%%%%%%%%%%%%%%%%%%%%%%%

In this paper we have addressed the problem of Hot Flasher,
investigating in details  the predicted evolutionary behavior of low
mass stars  with Z=0.0015 after an episode of mass loss during their
RG evolution. We found that stellar masses in the range $0.485 \le M
\le 0.497 M_{\odot}$ experience the He flash (and the explosive
H-reignition) during their WD cooling phase, when the residual H
shell burning has reduced the H-rich stellar envelope down to $M_e
\sim 0.0005 M_{\odot}$, independently of the mass of the model. Such
a result appears in reasonable agreement with theoretical
predictions given by Brown et al (2001) for the same  metallicity,
the small difference ($M \sim 0.0005$ against $0.0006 M_{\odot}$)
being likely the effect of small differences in the  adopted input
physics. Supporting, in turn, the evidence that the mass of the
H-rich envelope plays a critical role in the onset of the  delayed
He flashes.

According to this result, we are also predicting the structural
parameters needed to evaluate the  evolutionary times of structures
below the lower mass limit for   He ignition, which will definitely
cool down as He WD. We find that these He WD will reach the
luminosity $logL/L_{\odot}$=-1 in a time about 5 times longer than normal
Carbon-Oxigen WDs do, giving a detectable contribution to the
abundance of WD above such a luminosity if and when a not marginal
fraction of RG stars escape the He ignition.

For the sake of completeness, one has finally to advise that the
current scenario slightly depends on the luminosity of the RG models
undergoing the mass loss episode, since peeling off RG structures
either before (as in the previously  reported computations) or after
the first dredge up gives HF progenitors with different He
abundances in the stellar envelopes. Numerical experiments for our
$0.86 M_{\odot}$ models  have shown, e.g., that when the models is
stripped after the dredge up ($logL/L_{\odot}= 1.48, M_c=0.24M_{\odot}$)
the lower mass limit for LHF moves from 0.485 to 0.491 $M_{\odot}$, with  LHF
models characterized by slightly smaller H-rich envelopes, as given
by $M_e^f=0.00048 M_{\odot}$. As a whole, these appear as marginal
differences, not affecting the theoretical scenario we are dealing
with.

\section{Acknowledgments}

We warmly thank  Aldo Serenelli for  making available  models of
cooling dwarfs. It is also a pleasure to thank Santi Cassisi and
Giuseppe Bono for a critical reading of the manuscript and for
useful comments.

\end{document}